\documentclass{PoS}

\usepackage{bm}
\usepackage{amsmath}
\usepackage[square,comma,numbers,sort&compress]{natbib}
\usepackage{color}
\usepackage{graphicx}
\usepackage{bbm}

\newcommand*{\La}{\cal{L}}
\newcommand*{\no}{\noindent}
\newcommand*{\bea}{\begin{eqnarray}}
\newcommand*{\eea}{\end{eqnarray}}
\newcommand*{\be}{\begin{equation}}
\newcommand*{\ee}{\end{equation}}
\newcommand*{\pd}{\partial}
\newcommand*{\pdm}{\pd_{\mu}}
\newcommand*{\pdn}{\pd_{\nu}}

\newcommand*{\pref}[1]{(\ref{#1})}

\newcommand*{\mn}{{\mu\nu}}

\newcommand*{\nn}{\nonumber}

\title{Local and global gauge-fixing}

\ShortTitle{Local and global gauge-fixing}

\author{\speaker{Axel Maas}\thanks{Supported by the DFG under grant number MA 3935/5-1.}\\
        E-mail: \email{axelmaas@web.de}\\

        Institute of Theoretical Physics, Friedrich-Schiller-University Jena, Max-Wien-Platz 1, D-07743 Jena, Germany}

\abstract{Gauge-fixing as a sampling procedure of gauge copies provides a possibility to construct well-defined gauges also beyond perturbation theory. The implementation of such sampling strategies in lattice gauge theory is briefly outlined, and examples are given for non-perturbative extensions of the Landau gauge. An appropriate choice of sampling can also introduce non-trivial global symmetries as a remainder of the gauge symmetry. Some examples for this are also given, highlighting their particular advantages.}

\FullConference{Xth Quark Confinement and the Hadron Spectrum,\\
		October 8-12, 2012\\
		TUM Campus Garching, Munich, Germany}

\begin{document}

\section{Gauge-fixing as a sampling procedure}

Gauge theories have a very interesting structure. Take a theory of a set of fields $\phi$. If a fixed field value at each space-time point is given, this defines a configuration $\Phi\left[\phi\right]$ of the fields. In a gauge theory there exist local transformations $\phi\to\phi+\delta(x)$ of the fields $\phi$ such that for the set of configurations $\left\{\Phi\right\}$, called an orbit, reachable by these transformations the corresponding path integral remains invariant, i.\ e.\ all correlation functions have the same value. Especially, this implies that all correlation functions not invariant under a gauge transformation vanish. This vanishing is the realization of Elitzur's theorem \cite{Elitzur:1975im}. If this is the only symmetry of theory, the non-vanishing correlation functions define the non-trivial set of observables of the theory.

In principle, their calculation is all that is necessary to determine all experimental consequences of a theory. One possibility to do this are, e.\ g., lattice calculations. However, not all interesting cases can be solved with such methods efficiently. This has lead to the development of methods using a different approach, and which include, e.\ g., perturbation theory and functional methods.

In these cases, one removes the gauge symmetry, i.\ e., breaks it explicitly, in such a way as that any observable is not modified. This is done by instead of just integrating over the whole orbit in the path integral with a flat weight, integrating over it with some suitable non-flat weight: A non-trivial sampling of the orbit is introduced. This includes the extreme case of a $\delta$-function as weight to pick out a single element of the orbit, a single gauge copy. In general, if the sampling includes only a subset of the orbit, this subset will be called here the residual gauge orbit. If the residual orbit contains gauge copies which cannot be deformed into each other by infinitesimal gauge transformations, these are called Gribov copies \cite{Gribov:1977wm}.

For the implementation of this gauge-fixing procedure several possibilities exist. In lattice calculations it is indeed possible to perform this sampling explicitly \cite{Maas:2011se}. In perturbation theory, this can usually be achieved by the inclusion of auxiliary fields, the so-called ghost-fields \cite{Henneaux:1992ig}. The situation becomes complicated for non-lattice non-perturbative methods, like functional methods \cite{Maas:2011se,Maas:2011ba}. After the discussion of examples of such weighting procedures in section \ref{slattice}, the current state of the art in the continuum will be revised in section \ref{scont}. An interesting possibility is to include residual symmetries in the weighting process, which will be outlined in section \ref{sglobal}. Finally, a few concluding marks will highlight the current challenges in section \ref{ssum}. For the sake of being explicit, only Yang-Mills theory will be discussed here, though most of this can be found easily in other theories as well, e.\ g.\ QCD.

\section{Residual symmetries}\label{sglobal}

A starting point to illustrate the concept of residual symmetries are the covariant gauges in perturbation theory \cite{Bohm:2001yx}, described by the (Euclidean) path integral
\bea
<{\cal O}>&=&\int{\cal D}A_\mu{\cal D}c{\cal D}\bar{c} {\cal O}(A_\mu,c,\bar{c})e^{-\int d^4x {\La}_g}\label{pert}\\
{\La}_g&=&-\frac{1}{4}F_\mu^a F_{\mu a}+\frac{1}{2\xi}(\pdm A_\mu^a)^2+\bar{c}_a\pdm D_\mu^{ab} c_b\label{lagrangian}\\
F_\mn^a&=&\pdm A_\nu^a-\pdn A_\mu^a+g f^a_{bc} A_\mu^b A_\nu^c\nn\\
D_\mu^{ab}&=&\delta^{ab}\pdm+g f^{ab}_c A_\mu^c,\nn
\eea
\no where the gauge fields $A$ and the ghost and anti-ghost fields $c$ and $\bar{c}$ interact with a coupling strength $g$ and live in the adjoint representation of the gauge algebra codified by the structure constants $f^{abc}$, and $\xi$ is a parameter which modifies the sampling of the orbit with the Gaussian weight factor. The ghost field contribution only compensates, due to their origin from a Jacobian \cite{Bohm:2001yx}, the sampling in a way as to keep the values of the observables invariant. Thus, the last two terms of the Lagrangian \pref{lagrangian} represents the perturbative sampling procedure, the gauge-fixing. As a consequence, correlation functions not invariant under local gauge transformation no longer vanish.

These gauges have several continuous residual global symmetries, under which the sampling procedure is still invariant. One are global color rotations. As a consequence, any correlation function having a color direction still vanishes. To change this would require a further reduction of the sampled orbit. This is usually done when adding scalar fields in 't Hooft-type gauges \cite{Bohm:2001yx}, though this is not necessary \cite{Maas:2012ct}. Similarly, there is a trivial scale symmetry of the ghost fields, associated with ghost number conservation \cite{Bohm:2001yx}.

Far more interesting is the third symmetry, the BRST symmetry. This global symmetry transforms both ghost and gluon fields \cite{Bohm:2001yx}. Since it modifies the gluon fields without altering any gauge-invariant observables, this is necessarily a gauge transformation, though one which also alters the ghost fields. It thus connects the gauge copies, and the change in the ghost fields is necessary to transform the weight accordingly.

In the case of the perturbative Landau gauge, i.\ e.\ the limit $\xi\to 0$, the BRST symmetry can no longer act non-trivially on the gluon fields, as there is one and only one gauge copy per orbit which satisfies the perturbative Landau gauge condition $\pdm A_\mu^a=0$. However, there are still non-trivial ghost transformations, which leads to an additional ghost-anti-ghost symmetry \cite{Alkofer:2000wg}, essentially rotating ghost and anti-ghost into each other.

The situation becomes vastly more interesting when going non-perturbative. Then Gribov copies \cite{Gribov:1977wm,Singer:1978dk} appear. Thus BRST transformation can now act once more non-trivial even in Landau gauge, and one regains a non-trivial and non-perturbative BRST symmetry \cite{Maas:2012ct}. However, this requires, as it likely seems \cite{vonSmekal:2007ns,vonSmekal:2008es,vonSmekal:2008ws}, to include all Gribov copies with a flat weight. This leads to a number of complications, which can only be appreciated after investigating the appearance of Gribov copies a bit more closely.

\section{Landau-gauge examples on the lattice}\label{slattice}

Once non-perturbative, there are many Gribov copies satisfying the Landau gauge. The so obtained residual gauge orbit turns out to have a highly non-trivial structure \cite{Vandersickel:2012tg,Maas:2011se}. In particular, it is possible to define a bounded and convex Gribov region $\Omega$, enclosed by the so-called Gribov horizon, in which all eigenvalues of the Faddeev-Popov operator $-\pdm D_\mu^{ab}$ are strictly positive and only one vanishing on the horizon. Since every orbit passes through this region \cite{Dell'Antonio:1991xt}, it is possible to restrict the sampling of the orbit further to this region by the introduction of a $\Theta$-function on the Faddeev-Popov operator, with $\Theta(0)=1$.

Still, many Gribov copies remain inside this Gribov region. To deal with them, several different gauge conditions have been invented \cite{Maas:2011se}. These either attempt to identify a single Gribov copy as the representative of the gauge orbit, being the absolute Landau gauge \cite{Zwanziger:1993dh} and its inverse \cite{Silva:2004bv}, based on the so-called fundamental modular region of minimal gauge field norm or by externalizing either correlation functions \cite{Maas:2009se} or eigenvalues \cite{Sternbeck:2012mf} of the Faddeev-Popov operator. Since such constructions are notoriously complicated to construct in the continuum, this will not be discussed further here, see \cite{Maas:2011se} for more details.

It appears more interesting to pursue gauges averaging over the Gribov region with some prescription \cite{Maas:2011se,Maas:2011ba}. In general, any such averaging procedure is performed in the same way, i.\ e., by rewriting the perturbative expression \pref{pert} as \cite{Maas:2011se,Serreau:2012cg}
\be
<{\cal O}>=\lim_{\xi\to 0}\int{\cal D}A_\mu{\cal D}c{\cal D}\bar{c} {\cal O}(A_\mu,c,\bar{c})\Theta(-\pdm D_\mu^{ab})e^{-\int d^4x {\La}_g}w(A_\mu,c,\bar{c})\nn,
\ee
\no where $w$ is an appropriately chosen weight functions, which includes a normalization such that any observable remains unchanged. There are a few prominent examples in use.

In lattice calculations, a convenient choice is $w=1$, i.\ e.\ averaging over the first Gribov region with a flat weight. Since usual Landau-gauge-fixing algorithms appear to identify Gribov copies with equal probability, this is in practical calculations simplified to take a single, random representative for each configuration, leading to the minimal Landau gauge \cite{Maas:2011se}. In continuum calculation another choice is based on the (yet unproven) assumption that there exists a weight function such that $\Theta(-\pdm D_\mu^{ab})w(A_\mu,c,\bar{c})=\delta(-\pdm D_\mu^{ab})$ \cite{Vandersickel:2012tg}, i.\ e.\ averaging with a flat weight over the horizon only. This choice has the advantage that it can be rewritten as a local Lagrangian using further auxiliary ghost fields.

In contrast to these are gauges which average with some weight, either exponential \cite{Maas:2011ba,Maas:2011se} or Gaussian \cite{Serreau:2012cg}, with some argument over the Gribov region. Of course, once appropriately normalized, this does not change observables. Such an averaging includes a control parameter $\lambda$, essentially the width of the sampling function. This is a second gauge parameter, besides the perturbative one $\xi$.

To be concrete, consider the following two possibilities \cite{Maas:2011se,Serreau:2012cg}
\bea
w_1&=&\exp\left({\cal N}_1+\frac{\lambda_1}{V}\int d^dxd^dy\pdm^x\bar{c}^a(x)\pdm^yc^a(y)\right)\label{w1}\\
w_2&=&\exp\left({\cal N}_2-\frac{\lambda_2}{V}\int d^dx A_\mu^a A_\mu^a\right),\label{w2}
\eea
\no where the ${\cal N}_i$ are appropriately chosen normalizations. The first case \cite{Maas:2011se,Maas:2011ba} is a boundary term, essentially driving the ghost dressing function at zero momentum to a desired value with $\lambda_1$ being a Lagrange parameter. Especially, the value $\lambda_i=0$ is a fixed point returning the minimal Landau gauge, while the values $-\infty$ and $+\infty$ drive the ghost dressing function at zero momentum to the minimum and maximum possible values. In the second case \cite{Serreau:2012cg}, the Gribov copies are weighted with the norm of the gauge field. When $\lambda_2\to\infty$, this will put all weight on the Gribov copies in the fundamental modular region, i.\ e.\ the region of least gauge field norm. Again, the fixed-point value $\lambda_2=0$ returns the minimal Landau gauge.

\begin{figure}
\includegraphics[width=\linewidth]{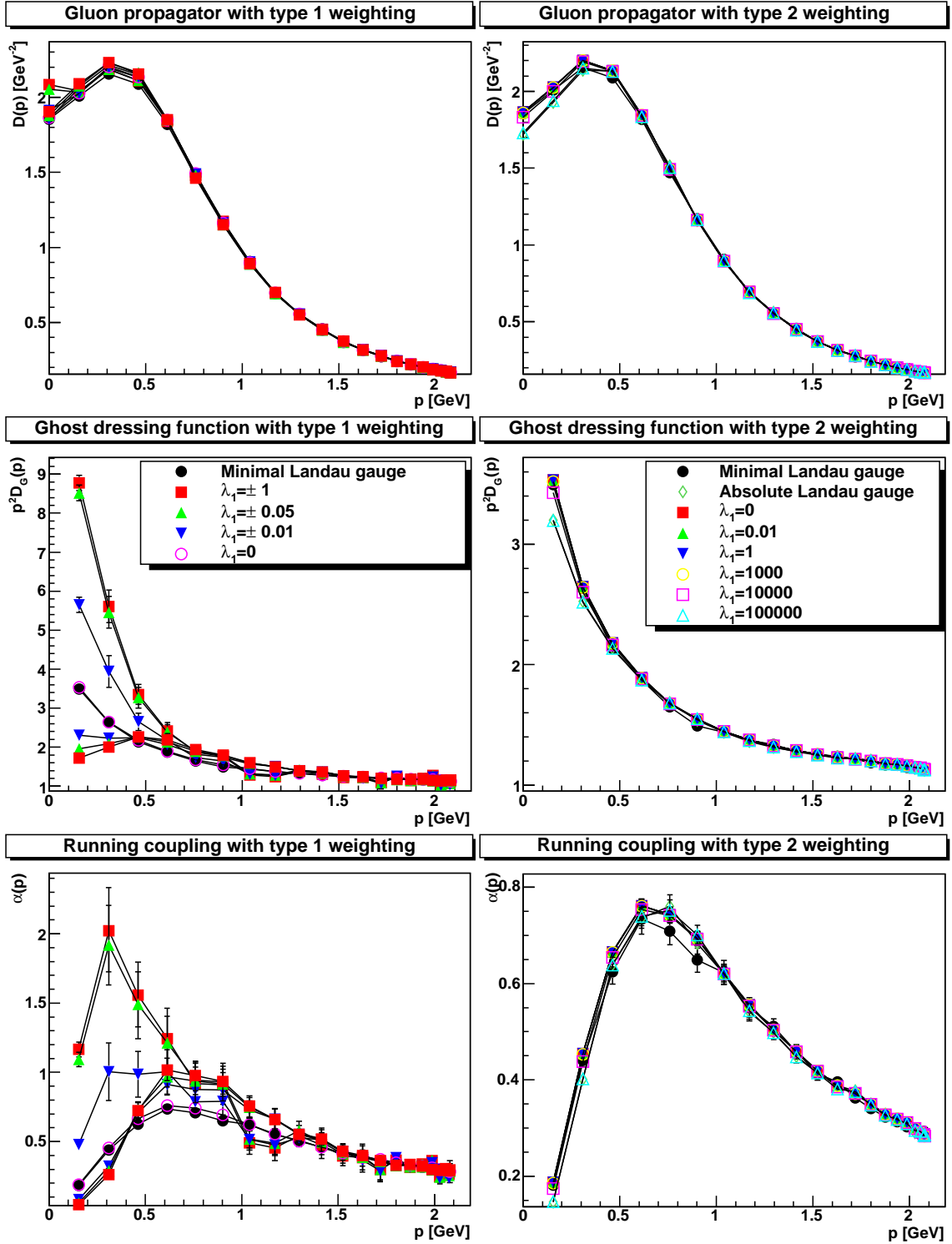}
\caption{\label{fig:3d}Preliminary results on the gluon propagator (top panels), ghost dressing function (middle panel), and running coupling (bottom panel) for the weight functions \protect\pref{w1} (left panels) and \protect\pref{w2} (right panels) for various values of the gauge parameters $\lambda_i$, compared to both the minimal Landau gauge and the absolute Landau gauge. Results are from three dimensions with $V=(7.9$ fm$)^3$ (42$^3$ lattice) at $a=0.189$ fm ($\beta=3.92$) \cite{Maas:unpublished}. Note the caveats of \cite{Maas:2011ba} on the problem of finding Gribov copies in lattice calculations. More details will be available elsewhere \cite{Maas:unpublished}.}
\end{figure}

To visualize the impact of both prescriptions, the gluon propagator, the ghost dressing function, and the running coupling in three dimensions are shown for several values of the $\lambda$s in figure \ref{fig:3d} \cite{Maas:unpublished}. First of all, it is clearly visible how, within statistical errors, the cases for $\lambda_i=0$ coincide with the minimal Landau gauge, as expected from the general arguments.

Concerning the weight function \pref{w1}, there are several observations, in agreement with earlier results \cite{Maas:2011ba}. When changing the gauge parameter $\lambda_i$ away from zero towards large positive and negative values, the results move towards those of the extremal maxB and minB gauges developed in \cite{Maas:2009se}, respectively. In these single-copy gauges the ghost propagator was extremalized. This is in agreement with the results and reasoning of \cite{Maas:2011ba}. In contrast to original hopes \cite{Maas:2009se}, but in agreement with insights based on the BRST construction in \cite{Fischer:2008uz,Maas:2010wb,Maas:2011se}, no qualitative change is observed in the infrared region. Especially the running coupling, though within large errors, remains vanishing. However, this will require more detailed investigations in the future for a more definite statement \cite{Maas:unpublished}, though any change would be surprising.

The situation for the weight function \pref{w2} is similar. Indeed, as argued \cite{Serreau:2012cg}, the results tend to agree better with the absolute Landau gauge with an increasing value of the gauge parameter $\lambda_2$, though the dependence is rather weak. Nonetheless, also this gauge fulfills in its lattice implementation its intended purpose.

However, in both cases it should be kept in mind that not all Gribov copies are guaranteed to be found in lattice simulations \cite{Maas:2011ba}. Thus the observe results are something of a lower limit to what the impact of Gribov copies really is. Thus an additional systematic uncertainty remains. Especially, it is never possible using the weighting function \pref{w2} in a lattice calculation using any contemporary algorithm to achieve a better approximation of the absolute Landau gauge than by any other method \cite{Maas:2011se,Maas:2008ri,Bogolubsky:2005wf,Cucchieri:1997dx,Silva:2004bv} of searching the absolute minimum, since it can never be better than the best Gribov copy found. But this does not exclude the possibility that new algorithms based on this idea maybe developed which can improve this.

\section{Taking the continuum limit}\label{scont}

Besides the technical problem of finding all Gribov copies, the problem remains of how to incorporate weights like \pref{w1} and \pref{w2} into continuum calculations. The $\theta$-function is known not to change the form of functional equations \cite{Zwanziger:2003cf}. This implies \cite{Maas:2011ba} that these equations contain the solutions both inside, outside, and in the whole of gauge field configuration space. These appear to consequently yield more solutions than observed on the lattice in the first Gribov region \cite{Boucaud:2008ji,Fischer:2008uz}. It remains thus an unsolved problem how to classify the continuum solutions of functional methods according to whether they belong to the first Gribov region or not, otherwise than by the (necessarily approximate) comparison to lattice results. Even the approach of \cite{Vandersickel:2012tg} cannot guarantee this, as it has not been ruled out that a similar replacement is justified from outside the first Gribov region.

Concerning the further weight \pref{w1}, the weight function is actually a surface term. Such terms turn into boundary conditions \cite{Maas:2011se,Garcia:1996np} in functional equations, which indeed appear to yield precisely the same results as the use of the weight function in lattice calculations \cite{Fischer:2008uz,Maas:2011se}. However, a formal understanding, or even proof, is still lacking.

The second weight \pref{w2} has so far only been used in the context of a perturbative calculations \cite{Serreau:2012cg}, similar in spirit to the approach of \cite{Vandersickel:2012tg}. Since the term is not a boundary term, it is not equivalent to a genuine boundary condition, though some arguments exist \cite{Maas:2008ri,Maas:2009se}, how this may be achieved indirectly.

In total, it remains still an open challenge how to formally correctly implement the same gauge condition on the lattice and in the continuum, though the rather good quantitative agreement obtained between both approaches \cite{Fischer:2008uz,Maas:2011ba,Maas:2011se}, as well as to other results, is already quite encouraging.

\section{Concluding remarks}\label{ssum}

In summary, it becomes clear that the structure of the gauge orbit and how to sample it, is irrelevant for any observable. This is, as it must be: The introduction of the gauge symmetry is only a technical tool to obtain a local field theory, which otherwise can only be described in terms of non-local objects, e.\ g.\ Wilson lines on a finite lattice. This also implies that any valid way of sampling the gauge orbit is equally admissible, and it is valid as long as it leaves observables invariant.

While this is a well-developed subject in perturbation theory \cite{Henneaux:1992ig}, it only becomes now more and more relevant beyond perturbation theory. The reason is simple. Various non-perturbative methods have reached a degree of maturity which makes them very capable of investigating physics. However, many of them, especially in the continuum, use a gauge-fixed framework, just like perturbation theory. Furthermore, all non-perturbative methods employ some kind of approximation, which cannot be systematically controlled. E.\ g.\ on a lattice this maybe the continuum limit, which is usually only addressed by extrapolations, and in continuum methods it may be the truncation of equation hierarchies \cite{Maas:2011se}. Thus, it is desirable to compare the results of different methods at the most elementary steps, to improve systematic reliability. Since these are often now gauge-fixed quantities, it is necessary to guarantee for a comparison that gauge orbits are sampled in the same way, since otherwise there maybe fundamental differences between the results. This has led historically to a number of problems \cite{Alkofer:2000wg,Maas:2011se}.

Besides the problems induced in this way by the sampling procedure, there is also an advantage. Choosing a suitable sampling may make problems tractable or simpler \cite{Maas:2011se,Maas:2012ct}. This has been widely used in perturbation theory \cite{Bohm:2001yx}. Harnessing this possibility in non-perturbative calculations, e.\ g.\ by formulations like \pref{w1} and \pref{w2}, will possibly help to solve many non-perturbative problems more efficiently.

\bibliographystyle{bibstyle}
\bibliography{bib}

\end{document}